\documentclass[aps,showpacs,preprint,superscriptaddress]{revtex4}

\usepackage{amsmath,amssymb,amsfonts}
\usepackage[english]{babel}
\usepackage{graphicx}

\begin{document}

\title{Optical Production of the Husimi Function of Two
Gaussian Functions}
\author{ J.R. Moya-Cessa, L.R. Berriel-Valdos and H.M. Moya-Cessa}
\affiliation{Instituto Nacional de Astrof\'isica \'Optica y Electr\'onica\\
Calle Luis Enrique Erro No. 1, Sta. Ma. Tonantzintla, Pue. CP 72840, Mexico}

\begin{abstract}
The intensity distribution of the Husimi function (HF)
and the squared modulus of the Wigner function (WF) are detected
in the phase space of an astigmatic optical processor. These
results, obtained in the laboratory, are compared against
numerical results generated by using analytical calculation for
the HF and WF. The signal function is the superposition of two
Gaussian functions with a separation between them, having the same
amplitude but a different variance.
\end{abstract}
\maketitle

\section{Introduction}
In 1932 Wigner introduced a distribution function in the context
of quantum mechanical correction to thermodynamic equilibrium
 \cite{wigner}. It is a bilinear function and its importance is due
to the fact that the WF describes a signal in the time and
frequency domains, called phase space, simultaneously
\cite{Roman-Moreno}. The WF can describe signals having two or
more variables, or their Fourier transforms. The WF of a signal
$\psi$(q), $W_{\psi}(q,p)$, has the important property that its
marginal probability in each coordinate is given by the
integration of $W_{\psi}(q,p)$ in its conjugate coordinate, i. e.,
$P(q)=\int{W_{\psi}(q,p)dp}$, and $P(p)=\int{W_{\psi}(q,p)dq}$. At
present, the WF has applications in many fields of physics and
engineering: in quantum mechanics it is useful for identifying
non-classic states, such as Fock states; in optics, to describe a
signal in the space and in the spacial frequencies simultaneously
so as in optical information processing \cite{Bastiaans}. This is
because the WF of a deterministic signal (totally coherent light)
relates Fourier optics and geometrical optics, and the Wigner
function of a stochastic signal (partially coherent light) relates
radiometry and partial coherence light theory.
\begin{figure}[h]
\centering
\includegraphics[width=0.8\textwidth]{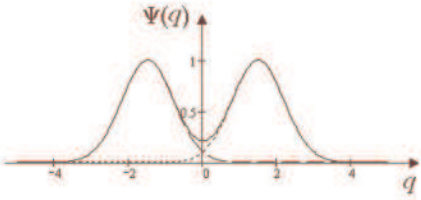}
  \caption{The sum of two Gaussian functions having the
same amplitude. They having a difference in variance and a
separation $q_{0}=1.5$.}
\end{figure}

Being the WF one of many distribution functions \cite{Bastiaans},
there are still some functions not introduced in the classical
(optics) world, namely the Glauber Sudarshan $P$-function
\cite{Glauber} and the Husimi $Q$-function
\cite{Husimi,Leonhardt}. It is therefore, the purpose of this
manuscript to present how the Husimi function (HF) may be obtained
in optics. The HF has a one-to-one correspondence with the WF and
to the optical signal (density matrix in quantum mechanics)
\cite{Moya}. One of the important properties of the HF is that its
distribution in the space is a radiometric observable that can be
measured directly. In quantum mechanics, the HF is refereed as a
classical quasi-probability distribution since it is a real and
non-negative quantity. In classical optics, it is directly
proportional to the intensity distribution detected in a
one-dimensional Fourier optical transformer, i.e., the HF of a
signal is the squared modulus of the Fourier transform of the
signal function times a weighted Gaussian function.

Here we will obtain the Husimi and (squared) Wigner functions for
two Gaussian functions \cite{Buzek} having different variances and
positions and will compare them against experimental results. The results may be easily extended to other kind of signals such as squeezed states \cite{vidiella,proceedings}. The
optical detection of these bilinear functions in the laboratory
was made using an astigmatic processor.

\section{The Wigner distribution function}

From its definition, we have that the Wigner function of a signal
is a 1D Fourier transform. Then, the optical description is in the
focal plane, called Fourier plane, of a cylindrical lens when used
in a 1D astigmatic transform system illuminated with a collimated
beam. Because it is detected the intensity of  light, in the
Fourier plane we have the squared absolute values of the Fourier
transform.
\begin{figure}[h]
\centering
\includegraphics[width=0.8\textwidth]{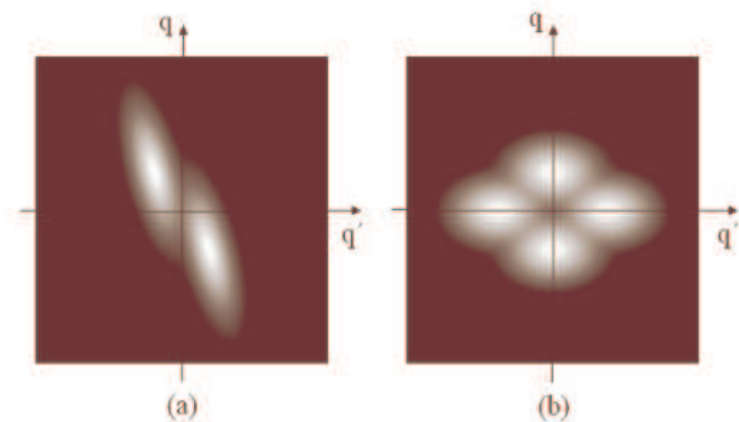}
  \caption{Bilinear functions $\Psi´(q,q')$ and
$\Psi_{W}(q,q')$. They were calculated numerically.}
\end{figure}
The Wigner distribution function in the space and
spatial-frequency domain can be written as (we use Dirac notation,
see \cite{eguibar})
\begin{equation}
W_{\hat\rho}(q,p)=\int^{\infty}_{-\infty}\exp(ipq')\langle{q-\frac{q'}{2}|\hat\rho|q+\frac{q'}{2}\rangle}dq'
\end{equation}
where q is the spatial variable and p is the spatial-frequency,
and $\hat\rho$ is the matrix of density. For an spatial signal
$\psi$(q) we have
\begin{equation}
\int^{\infty}_{-\infty}\exp(ipq')\langle{q-\frac{q'}{2}|\hat\rho|q+
\frac{q'}{2}\rangle}dq'=\int^{\infty}_{-\infty}\exp(ipq')\psi(q-\frac{q'}{2})\psi^{*}(q+\frac{q'}{2})dq'
\end{equation}
where the symbol * means  conjugation:
$\langle{q|\psi\rangle}=\psi(q)$ and
$\langle{\psi|q\rangle}=\psi^{*}(q)$ with $\psi(q)$ the  signal.
The density operator, $\hat\rho$, may be given in general by
\begin{equation}
\hat\rho=\sum_{n}\rho_{n}|\psi_{n}\rangle\langle\psi_{n}|
\end{equation}
i.e., a sum of different states.  From Eq. (2.2) we can see that
the Wigner function is the Fourier transform of the function
$\psi(q-\frac{q'}{2})\psi^{*}(q+\frac{q'}{2})$ having the
variables q and p as the canonic variables. In physics they
represent position and momentum but in optics they are associated
with position and spatial frequency.

The functions associated with the variables q and p are related by
\begin{equation}
\Psi(p)=F[\psi(q)]\qquad and \qquad \psi(q)=F^{-1}[\Psi(p)],
\end{equation}
where $F[.]$ means the Fourier transform and $F^{-1}[.]$ means the
inverse Fourier transform. Given a signal,$\psi(q)$, and its
Fourier transform, $\Psi(p)$, we have
\begin{equation}
W_{\psi}(q,p)=W_{\Psi}(q,p).
\end{equation}
From Equations (2.2) and (2.4) we have that $W_{\psi}(q,p)$ is the
Fourier transform, in the variable $q'$, of
\begin{equation}
r(q;q')=\psi(q+\frac{q'}{2})\psi^{*}(q-\frac{q'}{2})
\end{equation}
or, equivalently, $W_{\Psi}(q,p)$ is the Fourier transform, in the
variable $p'$, of
\begin{equation}
R(p;p')=\Psi(p+\frac{p'}{2})\Psi^{*}(p-\frac{p'}{2}),
\end{equation}
where $r(q;q')$ and $R(p;p')$ are related through a 2D Fourier
transform. So, the WF can only be reconstructed by a holographic
process and reconstructed holographically. Through an astigmatic
optical processor that can be detected is the squared modulus of
$W_{\psi}(q,p)$ or $W_{\Psi}(q,p)$.

\section{The Husimi distribution function}
The HF of a given signal is defined as
\begin{equation}
Q_{\psi}(q,p)=\frac{1}{\pi}\langle\alpha|\rho|\alpha\rangle,
\end{equation}
where $|\alpha\rangle$ is a so-called coherent state
\cite{Leonhardt}. The above equation, written in normal notation
is
\begin{equation}
Q_{\psi}(q,p)=\frac{1}{\pi}|\int_{-\infty}^{\infty}\psi_{\alpha}(q,p;q')\psi(q')dq'|^{2},
\end{equation}
where  \cite{Leonhardt} $\psi_{\alpha}(q,p;q')$ is given by
\begin{equation}
\psi_{\alpha}(q,p;q')=\pi^{-\frac{1}{4}}exp[-\frac{(q'-q)^2}{2}+ip(q'-\frac{q}{2})],
\end{equation}
which can directly be detected through an astigmatic optical
processor since it is the squared modulus of a Fourier transform.
The term $exp(-ipq')$ in Eq.(3.3) being the kernel of the Fourier
transform and the function to be transformed is
\begin{equation}
\psi´(q,q')=\sqrt[4]{\pi}\psi(q')\exp[-\frac{(q'-q)^{2}}{2}],
\end{equation}
that is the equivalent expression of $r_{\psi}(q,q')$ given in Eq.
(2.6) for the Wigner function.

\subsection{The Wigner function}
The signal considered in this work is two Gaussian functions
having unitary amplitudes and centered a quantity $q_{0}$ and
$-q_{0}$ from the origin
\begin{equation}
\psi(q)=\exp[-b(q-q_{0})^{2}]+\exp[-(q+q_{0})^{2}]
\end{equation}
where b is the inverse of the relative standard variation, i.e.,
the standard variation of the first Gaussian function divided by
that of the second Gaussian function. Fig. 1.1 shows the function
$\psi(q)$ for $q_{0}=1.5$ and b=1 that it has a minimum at origin.
By substituting Eq. (5) in (2) we have
\begin{eqnarray}
W_{\psi}(q,p)&=&\sqrt{\frac{2\pi}{b}}\exp[-\frac{4b^{2}(q-q_{0})^{2}+
p^{2}}{2b}]+\sqrt{2\pi}\exp[-\frac{4(q+q_{0})^{2}+p^{2}}{2}\nonumber\\&+&
4\sqrt{\frac{\pi}{b+1}}\exp[-\frac{4b^{2}q^{2}+
p^{2}}{b+1}]\cos{[2p(\frac{b-1}{b+1}q+q_{0})]}.
\end{eqnarray}

The first and the second terms of this equation are the Wigner
functions of the each terms of Eq. (12), respectively. The third
term is the interference term, which gives the negative values in
the Wigner function. The intensity distribution in the phase space
is then
\begin{equation}
I_{W}(q,p)=W_{\psi}^{2}(q,p).
\end{equation}

\subsection{The Husimi function}
For the Husimi function, by substituting Eqs. (3.3) and (3.5) in
Eq. (3.2), we have
\begin{eqnarray}
Q_{\psi}(q,p)&=&\frac{2}{\sqrt{\pi}}[\frac{1}{2b+1}\exp[-\frac{2b(q-q_{0})^{2}
+p^{2}}{2b+1}]+\frac{1}{3}\exp[-\frac{2(q+q_{0})^{2} +p^{2}}{3}]\nonumber\\
&+&\frac{2}{\sqrt{3(2b+1)}}\exp[-\frac{1}{3}\frac{(b-1)(q-q_{0})^{2}+2(2b+1)(q^{2}+
q_{0}^{2})+(b+2)p^{2}}{2b+1}]\nonumber\\&\cdot&\cos[\frac{2p}{3}\frac{(b-1)q-(5b+1)q_{0}}{2b+1}]].
\end{eqnarray}
\begin{figure}[h]
\centering
\includegraphics[width=0.8\textwidth]{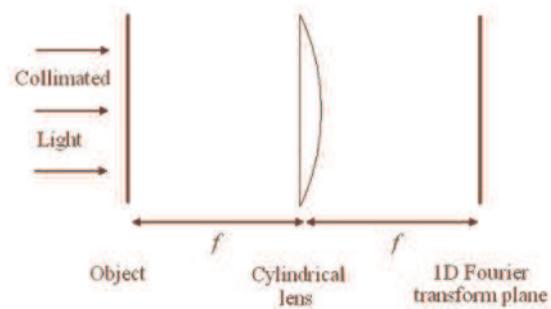}
  \caption{The astigmatic processor used in this work.
A cylindrical lens is illuminated with a collimated beam.}
\end{figure}
\begin{figure}[h]
\centering
\includegraphics[width=0.8\textwidth]{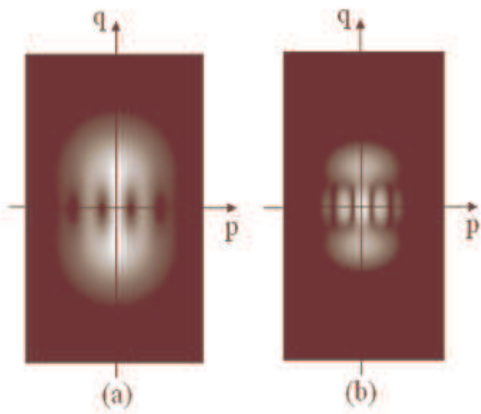}
  \caption{Distributions in gray levels of (a) the
Husimi function, and (b) squared modulus of the Wigner function.}
\end{figure}
\begin{figure}[h]
\centering
\includegraphics[width=0.3\textwidth]{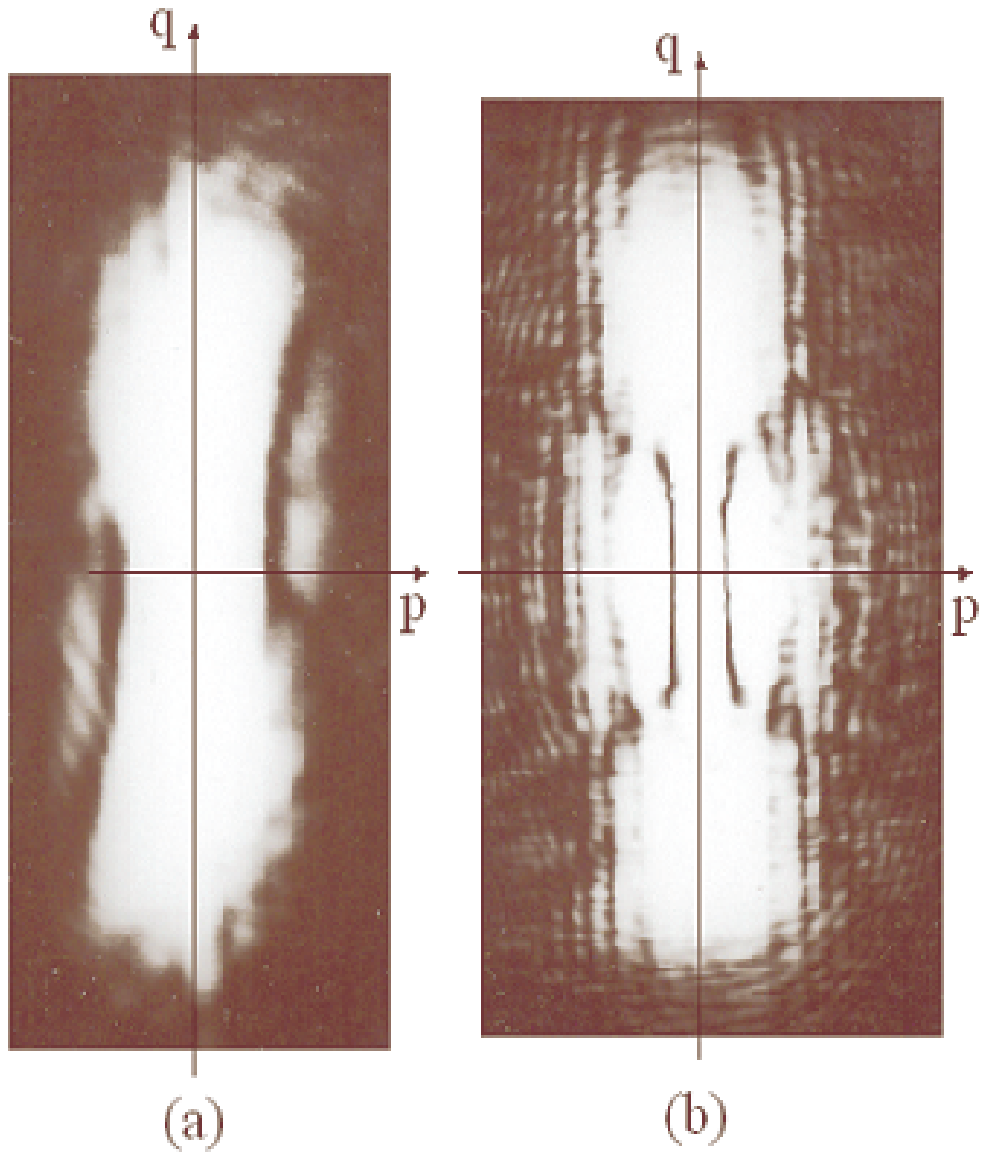}
  \caption{Images obtained in laboratory. (a) For
Husimi function, and (b) for the squared modulus of the Wigner
function. These have to be compared against those given in Fig.
3.4.}
\end{figure}

Figs. 2.2a and 2.2b show the gray level plots of the numerical
values obtained for the bilinear signals given in Eqs. (3.4) and
(2.6), respectively, for the signal given in Eq. (3.3). These
plots were taken like the objects to be used in the astigmatic
optical processor.

\section{Experimental Process}
Fig. 3.3 is an schematic diagram of a 1D Fourier transform
processor, the astigmatic optical processor used for detecting the
HF and the squared modulus of the WF. A convergent cylindric lens
is illuminated by a coherent collimated light, He-Ne laser beam
having a wavelength of 632.8 nm. In front of the lens is
collocated the object, a transparency  of Fig. 2.2a for the Husimi
function case and Fig. 2.2b for the Wigner function case. These
transparencies were photo-reduced $170$ times from its original
size. The negatives were made in a Technical Pan photographic film
from Kodak and developed using a D-19 dilution. In order to avoid
the vignetting effects, the transparencies were put in contact
with the cylindrical lens   \cite{Goodman}. Because of  the
photographic process some noise sources are present. The printing
process and the non linearity response of the film reduce and
change the gray tones of the photoreduction. In the optical
reproduction, another sources of errors are present as the spatial
coherence of the light source and the aberrations of the lens that
produces strong speckle.

 Fig. 3.4 shows the results obtained
for the Husimi and Wigner functions, Fig. 3.4a and 3.4b
respectively. A mechanical misalignment of the photoreduction film
produces the effect that can be seen in Fig. 3.5a were the
secondary lobes, i.e. the maxima of intensity are moved down and
up from their nominal positions. A good alignment produces a good
reconstruction, as can be seen in Fig. 3.5b.

\section {Conclusions}
We have detected in gray levels the Husimi function and the
squared modulus of the Wigner function. The function used as a
signal to be recorded is the superposition of two Gaussian
functions displaced from each other a quantity $2q_{0}$ and having
two different standard variances. They are compared against those
obtained using theoretical calculation, which were plotted
numerically. In spite of many sources of error are present during
the recording, the results given are good enough for recognizing
the HF and the WF of the signal. Finally, we have introduce in
classical optics a commonly used distribution function in quantum
mechanics, namely, the Husimi $Q$-function.

\end{document}